\documentclass[12pt,preprint]{aastex}
\usepackage{amsmath}

\begin{document}

\title{Magnetic Structure and Dynamics of the Erupting Solar Polar Crown Prominence on 2012 March 12}

\author{Yingna Su\altaffilmark{1,2} , Adriaan van Ballegooijen\altaffilmark{2}, Patrick I. McCauley\altaffilmark{2}, Haisheng Ji\altaffilmark{1}, Katharine K. Reeves\altaffilmark{2}, 
Edward E. DeLuca\altaffilmark{2}}
\altaffiltext{1}{Key Laboratory for Dark Matter and Space Science, Purple Mountain Observatory, Chinese Academy of Sciences, Nanjing 210008, China}  
\altaffiltext{2}{Harvard-Smithsonian Center for Astrophysics, Cambridge, MA 02138, USA.}  
\email{ynsu@pmo.ac.cn}

\begin{abstract}

We present an investigation of the polar crown prominence that erupted on 2012 March 12. This prominence is observed at the southeast limb by SDO/AIA (end-on view) and displays a quasi vertical-thread structure. Bright U-shape/horn-like structure is observed surrounding the upper portion of the prominence at 171~\AA~before the eruption and becomes more prominent during the eruption. The disk view of STEREO$\_$B shows that this long prominence is composed of a series of vertical threads and displays a half loop-like structure during the eruption. We focus on the magnetic support of the prominence vertical threads by studying the structure and dynamics of the prominence before and during the eruption using observations from SDO and STEREO$\_$B.  We also construct a series of magnetic field models (sheared arcade model, twisted flux rope model,  and unstable model with hyperbolic flux tube (HFT)). Various observational characteristics appear to be in favor of  the twisted flux rope model. We find that the flux rope supporting the prominence enters the regime of torus instability at the onset of the fast rise phase, and signatures of reconnection (post-eruption arcade, new U-shape structure, rising blobs) appear about one hour later. During the eruption, AIA observes dark ribbons seen in absorption at 171~\AA~corresponding to the bright ribbons shown at 304~\AA, which might be caused by the erupting filament material falling back along the newly reconfigured magnetic fields. Brightenings at the inner edge of the erupting prominence arcade are also observed in all AIA EUV channels, which might be caused by the heating due to energy released from reconnection below the rising prominence.

\end{abstract}

\clearpage


\section{INTRODUCTION}

The solar corona contains sheared or twisted magnetic fields overlying
polarity inversion lines (PIL) on the photosphere. The sheared/twisted fields can be
observed as filament channels on the disk and as coronal cavities in
limb observations; solar prominences are located within these regions. 
These structures warrant investigation because
of their role in prominence eruptions, coronal mass ejections (CMEs),
and solar flares. Understanding the topology and evolution of the prominence/cavity magnetic field structure
as well as the thermodynamics of the plasma within and surrounding prominences prior
to the eruption is key to understanding the initiation of solar eruptions.
 
Solar prominences are relatively cool structures embedded in the
million-degree corona. In H$\alpha$ when viewed above the solar limb, prominences appear as bright structures against the
dark background, but when viewed as ``filaments'' on the solar disk
they are darker than their surroundings. We will use the terms ``filament" and ``prominence" 
interchangeably in general. For more detailed reviews on observations and modeling of 
solar prominence, see \citet{1985SoPh..100..415H, 2010SSRv..151..243L, 2010SSRv..151..333M, 2014LRSP...11....1P, 2014IAUS..300..127V}. 
When observed on the solar disk with high spatial resolution, filaments show thin thread-like structures that continually evolve \citep{2008ASPC..383..235L}. Recent observations with the Solar Optical Telescope on the Hinode satellite and SDO/AIA have revolutionized our understanding of quiescent and intermediate prominences. When observed above the solar limb, such prominences always show many thin thread-like structures. In some cases the threads are mainly horizontal \citep[e.g.,][]{2007Sci...318.1577O}, in other cases they are mainly vertical \citep[e.g.,][and references therein]{2008ApJ...676L..89B, 2012ApJ...757..168S}. Hedgerow prominences consist of many thin vertical threads organized in a vertical sheet or curtain. Upward-moving plumes and bubbles have been observed in between the denser, downflowing threads \citep[e.g.,][]{2008ApJ...676L..89B}. Other prominences consist of isolated dark columns standing vertically above the PIL, and such prominences often exhibit rotational motions reminicent of ``tornados" in the Earth's atmosphere \citep[e.g.,][]{2012ApJ...752L..22L, 2012ApJ...756L..41S, 2013A&A...549A.105P}. The rotational motions have been confirmed using Doppler shift measurements \citep[e.g.,][]{1984SoPh...91..259L, 2012ApJ...761L..25O, 2014ApJ...785L...2S}.  Many quiescent prominences have horn-like extensions that protrude from the top of the spine into the cavity above \citep[e.g.,][]{2012ASPC..463..147B, 2013ApJ...770...35S, 2012ApJ...757..168S}. These horns may outline a flux rope located above the prominence \citep{2012ASPC..463..147B}.

The cool prominence plasma must somehow be supported against gravity because without such support the plasma would fall to the chromosphere on a time scale of about 10 minutes. It has long been assumed that prominences are threaded by horizontal magnetic fields and electric currents, which provide the upward Lorentz force needed to counter gravity \citep[e.g.,][]{1957ZA.....43...36K, 1974A&A....31..189K, 1983SoPh...88..219P}. The horizontal threads may be understood as plasma being supported at the dips of sheared arcade \citep{1994ApJ...420L..41A, 2002ApJ...567L..97A} or twisted flux rope \citep{1974A&A....31..189K, 2004ApJ...612..519V, 2006ApJ...641..590G}, but the structure of the vertical threads is not yet fully understood. For example, for tornado-like prominences it is unclear how horizontal fields can survive in the presence of rotational motions of the vertical structures \citep{1984SoPh...91..259L}. \citet{2010ApJ...711..164V} proposed that hedgerow prominences are located in vertical current sheets, and that the plasma is supported by small-scale ``tangled" magnetic fields within the sheet.  \citet{2012ASPC..463..147B} proposed that the current sheet is located below an elevated flux rope.  A three-dimensional MHD simulation performed by \citet{2012ApJ...758...60F} suggests that the vertical threads and horn-like structure can be explained by a magnetic flux rope with hyperbolic flux tube (HFT), which is the generalization of a 2.5D X-point or a 3D separator. The vortex-filament concept proposed by \citet{2014ApJ...785L...2S} is similar to the ``tangled field model" in the sense that the magnetic fields supporting the vertical threads are vertical and twisted. One problem with these ideas is that small-scale tangled fields are not consistent with the apparent smoothness of the magnetic fields derived from Hanle measurements \citep[e.g.,][]{2012ApJ...761L..25O}. 
 
Several mechanisms have been proposed to trigger solar eruptions. The initiation mechanisms of the explosive eruption can be 
divided into two categories: (1) ideal MHD instability or loss of equilibrium; (2) non-ideal fast magnetic reconnection.  The first group includes the kink instability\citep{2007ApJ...668.1232F}, torus instability \citep{2006PhRvL..96y5002K, 2014ApJ...789...46K} or loss of equilibrium \citep{1990JGR....9511919F, 2000JGR...105.2375L}; while the flux emergence model \citep{2000ApJ...545..524C}, flux cancelation model\citep{1989ApJ...343..971V}, and the breakout model \citep{1999ApJ...510..485A} belong to the second category.
At present, it is often unclear which mechanism, or combination of mechanisms, is responsible for any particular event.

The magnetic field plays a primary role in filament formation, stability, and eruption
\citep{Priest1989, 1995ASSL..199.....T, 2010SSRv..151..333M}. However, the 
magnetic structure of prominences is still not fully understood, with many observations
and theoretical models differing on the exact nature of the magnetic field. In this paper, we
will study an eruptive polar crown prominence consisting of vertical threads and 
examine the possibilities of three types of prominence support models, namely:
sheared arcade model, twisted flux rope model,  and tangled field model. The initiation mechanism 
of the explosive eruption will also be investigated.

\section{Observations}

\subsection{Data Sets and Instruments}

A large polar crown prominence was observed to erupt on March 11--12  in 2012, by the Atmospheric Imaging Assembly \citep[AIA, ][]{2012SoPh..275...17L} aboard the $\emph{Solar Dynamics Observatory}$ (SDO), as well as the STEREO$\_$B (Behind)/EUVI \citep{2004SPIE.5171..111W, 2008SSRv..136...67H}. At the time of eruption, STEREO$\_$B is observing the Sun ahead of the Earth, and the separation angle with Earth is 118$^{\circ}$. The photospheric magnetic field information is provided by the Helioseismic and Magnetic Imager \citep[HMI, ][]{2012SoPh..275..229S} aboard SDO.  

The apparent slow rise of the large prominence begins around 17:00 UT on 2012 March 11. This prominence eruption is associated with a CME, which has a median velocity of 399 km s$^{-1}$, according to the CACTus SOHO/LASCO CME catalog\footnote{http://sidc.oma.be/cactus/catalog.php}. The first appearance of the associated CME in the LASCO/C2 field of view is at 01:25 UT on 2012 March 12. Two bright ribbons and post-eruption arcade are observed during the eruption, while no associated flare can be clearly identified. In the current paper, we study the structure and dynamics of the prominence before and during the eruption.

\subsection{Structure and Dynamics of the Prominence-Cavity System before the Eruption}

Figure 1 presents multi-wavelength SDO/AIA observations of the target polar crown prominence system at 17:01 UT on 2012 March 11 prior to the eruption. 
This figure shows that the prominence is composed of a series of vertical threads. In 131~\AA, 94~\AA, and 304~\AA~images, the prominence displays dark absorption in the
middle and bright emission in the surroundings. The extended bright emission surrounding the dark filament is best seen at 171~\AA, which shows
that a bright U-shape/horn-like structure appears to be located above the dark vertical prominence threads.  In 335~\AA, 211~\AA, and 193~\AA~images, the prominence is seen in absorption, 
and the dark vertical threads are surrounded by a dark region in contrast to the bright emission shown at 171~\AA. At these three channels, we also see a cloud of bright emission (best seen in 211~\AA~images) located above the vertical threads within the dark cavity. Some of the prominence vertical threads can be seen in emission at 1600~\AA.

%This prominence appears to be similar to the tornado-like prominence studied by \citet{2012ApJ...756L..41S}.

AIA images at 304~\AA~(left), 171~\AA~(middle), and 193~\AA~(right) at 12:00 UT from March 6 to Mar 11 in 2012 are presented in Figures 2--3.  These two figures show that the height of prominence (304~\AA) appears to increase with time, and the size of the cavity (193~\AA) also appears to become bigger as time goes on. Due to the large time range, the prominence
at the limb is a projection of different portions of the long prominence. Therefore, the apparent increase of the prominence height and cavity size may be due to two reasons: the height of different parts of the prominence is different; or the whole prominence is rising with time.  The end view of the prominence on March 6 shows that the vertical threads are located at the bottom of the dark cavity (Figure 2c). A series of nearly vertical horn-like structure appears to go through the entire vertical threads from bottom to top on the southern side of the prominence in 171~\AA~image (Figure 2b). The side view of the prominence at 171~\AA~on March 7--8 shows nearly horizontal bright horn-like structure protruding the top of the vertical prominence threads. The cavity appears to become much bigger on March 8 (Figure 2i), although lots of bright loops along the line of sight are projected within the cavity. As the Sun rotates from March 9 to March 11, AIA observes the prominence from side view to end view (Figure 3). The vertical threads and horn-like structure are much wider on March 9 (Figures 3a--3b). Figures 3b--3c shows that  in 193~\AA~image bright U-shape structure appears at both the top and bottom parts of the vertical threads, while at the other part bright U-shape structure is shown at 171~\AA. This suggests that the U-shape structure goes through nearly the entire vertical threads when placing the two images together, and the same result can also be obtained from the observations on March 10--11. On March 11, the structure containing the vertical threads appears much narrower since the threads gradually align with each other along the line of sight, and the horn-like structure becomes more like a  narrow U-shape structure. The size of the cavity outlined by the big bright  loop is similar to that on March 8, though it is still filled with smaller bright loops projected along the line of sight.

\subsection{Structure and Dynamics of the Eruption}

SDO/AIA and STEREO$\_$B/EUVI observations of the prominence eruption  (before: top, during: middle, after: bottom) in  three wavelengths  (304~\AA: left, 171~\AA: middle, 193~\AA: right), are presented in Figures 4--5, respectively. Details of the eruption process can be found from the online videos 1 and 2. The end view by AIA in Figure 4 shows that at 00:30 UT on March 12 the prominence rises much higher and becomes much wider with two dark columns on the two sides and fewer dark thin threads in between. Small post-eruption loops are observed near the limb at 193~\AA~(Figure 4f). The prominence leaves the AIA field of view (FOV) at 04:00 UT, the 193~\AA~image shows much brighter and larger post-eruption loops (Figure 4i), and a helmet streamer structure appears in 171~\AA~image (Figure 4h).

The top/side view by EUVI$\_$B in Figure 5 shows that this prominence is very long, and the part where the prominence erupts first is near the western end, which appears to be higher than
the other part at 17:00 UT on March 11.  This eruption appears to be an asymmetric eruption, during which only the western leg lifts off, while the eastern part remains attached to the surface as shown in Figure 5d (also see video 2).  After the eruption, series of bright post-eruption loops are observed in the 195~\AA~images (e.g., Figure 5i).

The kinematics of the erupting prominence is presented in Figure 6. 
We trace the leading edge of the prominence to infer its bulk motion 
by first selecting the linear slice (black) that best characterizes the overall trajectory. 
Two additional slices (blue and green) offset by 2$^{\circ}$ in either direction are processed 
identically for error estimation (Figure 6a). Emission along 
these lines at a given time is binned to 300 pixels and interpolated onto a 
uniform distance grid, yielding a spatial resolution of $\sim$2$\arcsec$. 
Light curves are drawn from 10-frame (2-min) averages to improve 
signal-to-noise and are stacked against subsequent observations 
to produce height-time images like the one shown in Figure 6b
for the main slice (black). The height-time images are then further processed 
to improve contrast. Each row is multiplied by its height to boost signal 
far from the limb and the image is thresholded above 1.5$\times$ its median 
value. The Canny edge detection algorithm \citep{Canny1986} is then applied 
to extract the leading edge. Figure 6c shows the 
application of the Canny algorithm to the image in Figure 6b, 
and the pixels highlighted in red are used as the individual height measurements. 
These points are extracted automatically, but their time range must be selected 
manually, which is particularly important for the start time because it effectively 
defines the onset of the slow-rise phase. The aforementioned procedure is also used by \citet{Reeves2015} for IRIS 
observations of an eruptive prominence and by \citet{McCauley2015} for a 
statistical study that includes this event. 

The height measurements are then fit with an analytic approximation presented 
by \citet{2013ApJ...769L..25C} for their study of an active region flux rope eruption: 

\begin{equation}  \label{eq-fit}
h(t) = c_0 e^{(t-t_0)/\tau} + c_1 (t - t_0) + c_2
\end{equation}

\noindent $h(t)$ is height, $t$ is time, and $\tau$, $t_0$, $c_0$, $c_1$, and $c_2$ 
are free parameters. This model combines a linear equation to treat the slow-rise 
phase and an exponential to treat the fast-rise. The onset of the fast-rise phase can 
be defined as the point at which the exponential component of the velocity equals 
the linear (i.e., the total velocity equals twice the initial), which occurs at:

\begin{equation}  \label{eq-onset}
t_{\rm onset} = \tau \rm{ln}(c_1 \tau / c_0) + t_0
\end{equation}

Fitting is accomplished using MPFIT, a non-linear least squares curve fitting package 
for IDL \citep{2009ASPC..411..251M}. Figures 6d, 6e, and
Figure 6f show the fit result and its time derivatives. Based on this 
approximation, we find that the initial slow-rise velocity is 2.6 $\pm$ 0.2 km s$^{-1}$, 
and the maximum velocity in the AIA  FOV is 110 $\pm$ 5  km $s^{-1}$. The onset of the fast rise 
phase occurs at 22:57 UT $\pm$ 7 minutes at a height of 97 $\pm$ 5 Mm. At the onset 
point, the acceleration is 1.2 $\pm$ 0.1 m s$^{-2}$, and the final acceleration 
is 49 $\pm$ 5 m s$^{-2}$. 

Two strategies are employed to quantify uncertainty, and the errors quoted are the sum 
of both. The first is to identically process two adjacent slices offset by 
2$^{\circ}$ on either side of the original. Standard deviations from these results 
account for $\sim$55\% of the velocity, acceleration, and time uncertainties. 
The second strategy is to perform 100 Monte Carlo (MC) simulations to estimate uncertainties 
from the fit parameters by randomly varying our height measurements within some assumed 
error and refitting Equation~\ref{eq-fit}, as was done by \citet{2013ApJ...769L..25C}. 
Since there are no standard errors for height measurements 
obtained by our edge detection method, 
the assumed height errors are chosen to yield a reduced 
chi-squared ($\chi^{2}_{R}$) of 1.0 for the fit. That gives us an 
uncertainty of 4.1$\arcsec$, which corresponds to $\sim$7 AIA pixels and $\sim$2 pixels 
on the height-time image. This value is included in the onset height uncertainty, to which it 
contributes 65\%, while uncertainties from the 3 separate slices and MC realizations account 
for 15\% and 20\%, respectively. 

%We can also compare our results to SOHO/LASCO observations to both test our 
%trajectory and infer motion occurring between the AIA and LASCO/C2 FOVs. 
In order to test the accuracy of our computed trajectory, we calculate the expected arrival time of the CME in SOHO/LASCO/C2 using two different assumptions.
If we assume a constant acceleration of 49 m s$^{-2}$  
after leaving the AIA FOV, the prominence would arrive at the LASCO/C2 FOV at around 02:14 
UT on March 12th with a velocity of $\sim$280 km s$^{-1}$. If, instead, the prominence continues along 
our height-time fit, it would appear in C2 at 02:05 UT with a velocity of $\sim$420 km s$^{-1}$. 
The actual arrival time of the prominence at 2.5 R{$_{\odot}$} is between 02:00 and 02:12 UT, and 
the velocity listed in the CACTus CME catalog is 399 $\pm$ 65 km s$^{-1}$. The CACTus velocity 
is derived from linear fits to measurements across the entire CME from both the C2 and C3 detectors, 
which cover 2.5 to 30 R{$_{\odot}$} \citep{2004A&A...425.1097R, 2009ApJ...691.1222R}. 
The prominence may then have followed our projected trajectory until nearly 2.5 R{$_{\odot}$}, 
after which it transitioned to a constant velocity or decelerated somewhat at large heights, but this 
is speculative given the lack of observations between AIA and LASCO/C2.

\subsection{Signature of Reconnection: U-shape and Blobs}

Figure 7 shows rotated views of AIA images from 19:00 UT on March 11 to  00:39 UT March 12, and details can be found at corresponding online video 3. The first and second rows show radial-filtered and
running-difference images at 171~\AA, and following the same format AIA images at 193~\AA~are presented at the third and fourth rows.
The prominence  gradually rises up from 19:00 UT to 23:00 UT on March 11. A post-eruption arcade (blue arrow) begins to appear around 23:50 UT on March 11.
A newly formed U-shape structure (green arrow) begins to show up at the lower part of the rising vertical threads around 00:10 UT on March 12. 

Similar to Figure 7, Figure 8 shows rotated AIA images from 00:43 UT to 01:19 UT on March 12. Small bright blobs (white arrow) begin to appear at the bottom of the erupting prominence
vertical threads around 00:51 UT.  The blobs are best observed at 171~\AA, and the bigger and brighter blobs can also be seen at 193~\AA. These bright blobs then rise up into the less dense prominence region located between the two dark dense vertical threads.  The blobs appear to rise faster than the bulk motion of the erupting prominence (see video 3).

A detailed kinematics study of the rising blobs is presented in Figure 9. Figure 9a shows AIA 171~\AA~running-difference image at 01:20 UT on 2012 March 12. Distance-time plots of emission at 171~\AA, 193~\AA, 211~\AA~along the white slice marked in Figure 9a are presented in Figures 9b, 9c, 9d, respectively. We perform linear fits (marked as red lines) to three of the rising blobs for each channel, and the velocity for each rising blobs is presented in the figure. This figure confirms that the speed of these bright rising blobs is larger than that of the leading edge and bulk motion of the prominence. The rising blobs that appear earlier (95 km s$^{-1}$) are slower than those appear later (259 km s$^{-1}$). The appearance height of these blobs is increasing with time as shown in Figures 9b--9d. In addition, bright falling features are also observed starting around 01:10 UT (at 171~\AA) on March 12 as shown in Figure 9b. 

The concave up U-shape structure, rising blobs, and post-eruption arcade are all signatures of reconnection. The aforementioned observations suggest that magnetic reconnection begins around 23:50 UT on March 11 near the bottom of the vertical threads. The increase of the appearance height of bright blobs may be due to the increase in height of the reconnection point.  Intermittent plasmoids or blobs are generally explained by the tearing-mode instability of the thin current sheet where a series of magnetic islands are recurrently created during reconnections \citep{1963PhFl....6..459F, 2006Natur.443..553D}. Small plasmoids or magnetic islands have been found to flow along the current sheet either sunward or anti-sunward in both observations \citep[][and references therein]{2010ApJ...722..329S, 2013MNRAS.434.1309L, 2013ApJ...767..168L} and simulations  \citep[][and references therein]{2011ApJ...737...14S, 2012ApJ...760...81K}. The speed of rising blobs in our eruption is at the lower end of the anti-sunward blobs'  speed in the literature (100--1400 km s$^{-1}$). This is not surprising, since our event is a slow eruption that occurred at the polar crown.% More discussions on the comparison with Alfven speed and sound speed as those in the literature???

\subsection{ Darkenings  and Brightenings during the Eruption}

Figure 10 shows AIA observations of bright and dark ribbons during the eruption.  In 304~\AA~(Figure 10a) image, we can see two bright ribbons located at the footpoints of the post-eruption arcade, which can be seen most clearly at 193~\AA~(Figure 10c). Corresponding to these bright ribbons, two dark ribbons are seen in absorption at 171~\AA~(Figure 10b). Distribution of intensity versus time along a slice (from south west to north east) nearly perpendicular to the ribbons (white line in the top left corner image) at 304~\AA~and 171~\AA~are presented in Figures 10d--10e. Figure 10d shows that the two bright ribbons first  appear around 01:30 UT,  and the two ribbons then gradually move away from each other. The slope of the outer edge of the ribbons, i.e., the newly formed ribbons, shows that the southern ribbon appears to move faster. The southern dark ribbon seen in absorption at 171~\AA~first appears around 01:10 UT, and the appearance of the full northern dark ribbon is around 01:30 UT. At 171~\AA, the ribbons are only seen in absorption initially, but later on bright emission also shows up. Most of the northern ribbons are seen in absorption, while only the outer edge of the southern ribbon, i.e., the newly formed ribbons, are dark.  

The observational characteristics of dark ribbons at 171~\AA~suggests that they might be caused by the erupting filament material falling back along the newly reconfigured magnetic fields. The co-spatial brightenings at 304~\AA~might be partially caused by heating of the plasma due to the kinetic energy of falling filament material compressing the plasma \citep{2013ApJ...776L..12G}. In addition, thermal and/or non-thermal energy released from reconnection impacting the chromosphere might also contribute to the observed brightenings, as suggested by the lateral appearance of brightenings at 171~\AA.
 
 Figure 11 shows brightenings of prominence material observed by AIA (left: 211~\AA, middle: 304~\AA) and EUVI$\_$B (right: 304~\AA) during eruption. The top row shows images before the appearance of 
 the brightenings, and the brightenings are shown in the bottom row. Brightening B1 first appears around 01:15 UT, and brightening B2 shows up (01:19 UT) immediately after. Both brightenings are observed in all EUV channels, and B1 appears to be brighter than B2, especially in the hotter channels (e.g., 94~\AA, 335~\AA). The change of the prominence seen from absorption to emission (B1 in Figures 11a and 11d) suggests that the brightenings might be caused by heating. EUVI$\_$B observation shows that these brightenings occur at the inner edge of the erupting prominence. Therefore, the observed prominence brightenings are likely to be caused by heating due to energy released from reconnection below the rising prominence.
 
%Shibasaki (2002) proposes that plasma falling from high altitudes after filament eruptions can convert potential energy into thermal energy, providing the source of energy and mass supply for long-duration solar flare events in the long-lasting decay phase. The impact sites show clear evidence of brightening in the observed extreme ultraviolet wavelengths due to energy release.  Two plausible physical mechanisms for explaining the brightening are considered: heating of the plasma due to the kinetic energy of impacting material compressing the plasma, or reconnection between the magnetic field of low-lying loops and the field carried by the impacting material. 

%By analyzing the emission of the brightenings in several SDO/Atmospheric Imaging Assembly wavelengths, and comparing the kinetic energy of the impacting material (7.6 ? 1026Ð5.8 ? 1027 erg) to the radiative energy (Å1.9 ? 1025Ð2.5 ? 1026 erg), Gilbert 2013 find the dominant mechanism of energy release involved in the observed brightening is plasma compression. Other phenomena with a similar observational signature are sequential chromospheric brightenings (SCBs; Balasubramaniam et al. 2005; Pevtsov et al. 2007; Kirk et al. 2012). Kirk et al. (2012) propose that SCBs are caused by the chromospheric impact of accelerated plasma along newly reconfigured magnetic field lines. 

\section{Modeling}

\subsection{Flux Rope Insertion Method}

Our magnetic field models are constructed using the flux rope insertion method developed by \citet{2004ApJ...612..519V}. 
Detailed description of the methodology can be found in the literature \citep{2008ApJ...672.1209B, 2011ApJ...734...53S, 2012ApJ...757..168S},
and we describe the method briefly below. First, the potential field is computed from the high-resolution (HIRES) and global magnetic maps. Then, by appropriate modifications of the vector potentials a ``cavity" is created above the selected path, and a thin flux bundle (representing the axial flux of the flux rope ($\Phi_{axi}$)) is inserted into the cavity. Circular loops are added around the flux bundle to represent the poloidal flux of the flux rope ($F_{pol}$). The above field configuration is not in force-free equilibrium. So our next step is to use magneto-frictional relaxation to evolve the field toward a force-free state. This method is an iterative relaxation method \citep{2000ApJ...539..983V} specifically designed for use with vector potentials. Specifically, we solve the following equation:
\begin{equation}
\frac{\partial{\boldsymbol {A}}}{\partial t}=\eta_{0}\boldsymbol v \times\boldsymbol{B}-\eta_{2}\nabla\times\boldsymbol{B}+\frac{\boldsymbol {B}}{B^{2}}\nabla\cdot(\eta_{4}B^{2}\triangledown\alpha)+\nabla(\eta_{d}\nabla\cdot\boldsymbol{A}),
\end{equation}
where $\boldsymbol v$ is the plasma velocity, $\eta_{0}$, $\eta_{2}$, $\eta_{4}$, and $\eta_{d}$ are constants in space, and $\alpha\equiv\boldsymbol{j}\cdot\boldsymbol{B}/B^{2}$, where $\boldsymbol{j}=\nabla\times\boldsymbol{B}$. The velocity is given by 
\begin{equation}
\boldsymbol {v}=(\it{f}\boldsymbol{j}-\it{v_{1}}\boldsymbol{\hat{r}}\times\boldsymbol{B})\times\boldsymbol{B}/B^{2},
\end{equation}
where $\it{f}$ is the coefficient of magnetofriction, and $\it{v_{1}}$ describes the effects of buoyancy and pressure gradients in the photosphere \citep{2008ApJ...672.1209B}. Magnetofriction has the effect of expanding the flux rope until its magnetic pressure balances the magnetic tension applied by the surrounding potential arcade. Significant magnetic reconnection between the inserted flux rope and the ambient flux may occur during the relaxation process. Therefore, the end points of the flux rope in the relaxed model may be different from that in the original model.

\begin{table}
\tabletypesize{\scriptsize}
\begin{center}
\caption{Parameters of Relaxation Method.}
\begin{tabular}{rrrr}
\tableline\tableline
\multicolumn{4}{c}{Relaxation} \\
\cline{1-4}
Iteration & $\eta_{0}$ & $\eta_{2}$ & $\eta_{4}$ \\
\tableline
0--100 & 1 & 0 & 0  \\
100--1000 & 1 & 0 & 0.003  \\
1000--10000 & 1 & 0 & 0.001\\
10000--20000 & 1 & 0 & 0.0003 \\ 
20000--70000 & 1 & 0 & 0.0001\\ 
\cline{1-4}
70000--80000 & 1 & 0.003 & 0.01 \\ 
\tableline 
\end{tabular}
\end{center}
\end{table}

The lower boundary condition for the HIRES region is derived from line-of-sight (LOS) photospheric magnetograms obtained with the SDO/HMI. Since the prominence is observed near the east limb, we use magnetograms that are taken several days after the prominence eruption on March 12. We combine four magnetograms taken on 2012 March 16--19 (each at 18:11 UT) to construct a high-resolution map of the radial component B$_{r}$ of the magnetic field as a function of longitude and latitude at the lower boundary of the HIRES region. We also use a SDO/HMI synoptic map of B$_{r}$ to compute a low-resolution global potential field, which provides the side boundary conditions for the HIRES domain, and also allows us to trace field lines that pass through the side boundaries of the HIRES region.

The HIRES magnetic map is shown in Figure 12. Note that this region is at the polar crown, and the field has mixed polarity with dominantly positive polarity on the south side of the PIL and negative polarity on the north side. Based on this map alone, it is difficult to recognize exactly where the PIL is located.  Therefore we used the filament channel observed by STEREO$\_$B to locate the base of the prominence/filament on the magnetic map. The blue curve is the path along which the flux rope will be inserted into the model. At the two ends of the path (blue circles), the flux rope is anchored in the photosphere.  Because this region is very close to the South pole, strong numerical artifacts begin to appear at the side near the South pole during the magneto-frictional relaxation process. To reduce the artifacts, we adopt a special relaxation procedure. During the first 70000-iteration relaxation we move the HIRES region close to the disk center by setting the latitude of the center of the HIRES box to be zero. According to our previous experience we find that it normally takes about 70000-iteration relaxation for the magnetic fields in the quiescent polar crown filament system to approach a force-free state. Then we move the HIRES region back to the original location, then run 10000 more iterations to make the fields to interact and merge with the real global polar magnetic fields. After this  extra 10000-iteration relaxation, the magnetic fields in the HIRES region become adjusted to the real global polar fields in the surrounding low-resolution region. If we continue to relax the magnetic fields, numerical artifacts begin to appear.  The two criteria that we use to choose the specific iteration numbers are: minimizing numerical artifacts and keeping the HIRES region being adjusted to the original surrounding polar magnetic fields. The parameters used during the relaxation process are presented in Table 1. 
%We construct five models with different combinations of  axial and poloidal fluxes of the inserted flux rope. Table 2 shows the model parameters for the five models.

\subsection{Models versus Observations}
 
  \begin{deluxetable}{ccccc}
\tabletypesize{\scriptsize}
%\rotate
\tablewidth{0pc}
\tablecaption{Model Parameters for the Polar Crown Filament System. }
\tablehead{ \colhead{Model} & \colhead{$\Phi_{axi}$}  & \colhead{$F_{pol}$ } & $E_{free}$ & $E_{free}$/$E_{poten}$\\
\colhead{No.} & \colhead{($10^{20}$ Mx)} & \colhead{(10$^{10}$ Mx cm$^{-1}$)} & \colhead{(10$^{30}$ ergs)} & (\%) } 
\startdata 
1 & 2 & 0 &  1.98 & 17  \\ 
2 & 3 & 0   & 2.98 & 26 \\   
3  & 2 & 2 & 8.14 & 71\\   
4  & 3 & 3  & 16.2 &141 \\  
5  & 3 & 1   &  5.76 & 50  \\
\tableline
\enddata
\tablenotetext{a}{The potential field energy is 1.15$^{31}$ erg}
\end{deluxetable}
 
We construct five models with different combinations of  axial and poloidal fluxes of the inserted flux rope. In all cases the inserted flux ropes have sinistral orientation of the axial field.
Table 2 shows the model parameters of the five models. The initial inserted flux bundles of Models 1 and 2 are straight (untwisted) bundles, while for Models 3--5 the initial inserted flux bundels are twisted
flux ropes with nonzero poloidal flux (F$_{pol}$). As mentioned in \citet{2011ApJ...734...53S}, one constraint on the stability of the models is that the magnetic energy of the 
field after the relaxation should be less than that of the open field. To estimate the energy of the open field, 
we change the negative polarity fields of the HIRES region to positive polarity and then compute a potential field. The resulting
open-field model has a magnetic energy $E_{open} = 9.77 \times 10^{31}$ erg, whereas the standard potential field has energy of $E_{pot} = 1.15 \times 10^{31}$ erg. Therefore, the free energy of the open field is about $8.62 \times 10^{31}$ erg. This requires that the free energy of the flux-rope models be less than $8.62 \times 10^{31}$ erg. Note that these energies refer to the HIRES part of the computational domain only (not the whole Sun). All of the five models presented in this paper meet this criterion.

Figure 13 presents the current distribution in a vertical plane indicated by the yellow line (nearly perpendicular to the PIL) in Figure 12 for Models 1--4 after relaxing for 80000 iterations.
The strong current is concentrated in the white region. The black and white vectors refer to the magnetic vectors. This figure shows that Model 1 is close to a normal 
sheared arcade, while Model 2 appears to be a flux rope configuration with an X-point/HFT and sheared arcade located  below. Model 3 and Model 4 display an 
elevated twisted flux rope configuration.  

Figure 14 shows comparisons of Models 1--4 with SDO/AIA and STEREO$\_$B observations.
The colored curves in Figure 14 show field lines selected by clicking different points in the two dimensional 
plots shown in Figure 13 from the four models. The blue features
indicate dips in the field lines (depth of color increases with height). The red and green contours 
refer to the HMI photospheric flux distribution in the HIRES region. The background images in the row 1,
row 2, and row 3 are taken at 171~\AA, 193 ~\AA~by SDO/AIA, and at 171~\AA~by STEREO$\_$B, respectively.
We can see that the dips of the field lines in Model 1 is lower than the prominence observed by both SDO/AIA and  
STEREO$\_$B/EUVI. Moreover, the cavity in the model is much smaller than observed, and the model cannot
reproduce the observed U-shape/horn-like structure. Therefore, we think that the sheared arcade topology does
not match the observations. For Model 2, the height of the dips of field lines matches the AIA observations well, but is much lower than
the height of prominence seen in the STEREO$\_$B  observations, especially the part where it erupts first.  This model exhibits an X-point topology as suggested 
by the simulation in \citet{2012ApJ...758...60F}, and the location and height of the U-shape structure in the model match 
the SDO/AIA observations well, but not the STEREO$\_$B observation. Moreover, the size of the ``U" in the model is much smaller
than that of the observation. The height of the dips of field lines in Models 3 and 4 correspond very well with the
height of the observed prominence by AIA and EUVI, though 
a small offset towards the south in the models is identified when compared with AIA observations. 
There appears to be different sizes of cavities in the observations, since
the prominence is very long. The size of flux rope  in Model 3 appears to match the foreground cavity, 
while for Model 4 it appears to match the background larger cavity  better.

Figure 15 shows comparisons of the ``U" structure in Model 4 (right column) and Model 5 (left column) with that in observations. The color curves in
this figure are selected field lines that going through a vertical plane near the location of ``U" structure 
where the prominence firstly erupts (white arrows in Figures 15a--b). The bright ``U"  structure in both AIA and EUVI observations 
are indicated with the solid white arrows. Corresponding U-shape structure in the models are marked with dashed white arrow (light blue and pink lines). 
This figure shows that the location, size, and height of the U-shape structure in Model 4 match the observations well. While for  Model 5, 
the ``U" structure is much lower than that in the observations.

%The flux rope is held down by an overlying arcade (not shown) but is
%close to the limit of stability. The height and width of the flux rope
%are severely constrained by the need to produce a stable model in which
%the magnetic pressure of the flux rope is balanced by tension of the
%overlying arcade.

\section{Discussions}

\subsection{Onset of Fast Rise: Torus Instability}

In Section 2.3, we found that the prominence eruption begins with slow rise then evolve to fast rise, which is 
often seen in solar filament eruptions \citep{2005ApJ...630.1148S, 2013ApJ...769L..25C}. The important question is what causes the transition from slow rise to fast rise? In other words, what causes the onset of explosive fast rise phase? In this section, we investigate two possibilities: i.e., torus instability and magnetic reconnection.

%As the flux rope rises it enters a domain where the strength of the overlying potential field falls-off quickly with height. In this region, the upward motion of the flux rope cannot be %restrained any more by the magnetic tension of the arcade. Aulanier et al. (2010) showed that when the flux rope axis reaches the height at which the decay index of the arcade $n = -
%\partial{lnB}/\partial{lnz}$ is above a certain threshold, it becomes unstable. ATDD10 found that the threshold was equal to $n_{tresh} = 1.5$, in agreement with the theoretical calculation %of the torus instability \citet{2006PhRvL..96y5002K}.

The threshold of torus instability is given in terms of the decay index of the external poloidal field, at the position of the 
current channel, $n = -\partial{lnB}/\partial{lnz}>n_{cr}$. The canonical values of the critical decay index are 1.5 for a 
toroidal current channel \citep{1978mit..book.....B} and 1.0 for a straight current channel \citep{1978SoPh...59..115V}.
The critical index ($n_{cr}$) for torus instability ranges from 1.0 to 2.0 in theoretical calculations and numerical simulations
\citep{2006PhRvL..96y5002K, 2007ApJ...668.1232F, 2010ApJ...708..314A, 2010ApJ...718.1388D}. On the other hand, from 
 the measurement  of the height of a set of quiescent prominences, combined with potential field extrapolations, \citet{2001JGR...10625177F} 
found $n_{cr} \approx 1$.

Figure 16 shows distributions of the magnitude of horizontal components of the potential field (top) and the decay index (bottom) of
the horizontal components with height over the main PIL at two locations marked with a yellow line (middle) and a star sign (right) in Figure 12. 
The dashed and solid vertical lines refer to the height of the outer edge of the prominence and the center of the dark cavity at the onset of 
 fast rise phase, respectively. Due to the variety of the  supporting magnetic configuration, we take the first (used by \citet{2001JGR...10625177F}) 
 and second heights as the lower and upper limits of the apex of the flux rope magnetic axis, respectively. We find that at the onset of fast rise phase,
 the decay index n is $1\pm0.2$, which is very close to the critical value of a straight current channel for torus instability. STEREO observations
 suggest that this prominence begins to erupt near the west end rather than the middle. Figures 16 shows that the lower limit of the decay index near
 the west end is slightly larger than that in the middle at the onset of fast rise. We should note that the potential field model is based on the magnetograms taken
 several days later, which is only a close approximation. The magnetic fields at the time of  the eruption may be different due to disturbance of a prominence eruption
 nearby several hour earlier.  Therefore, we cannot exclude the possibility that the overlying field actually falls off more quickly with height as the the reason
 why the western portion of the filament erupts firstly.  
 
\subsection{Magnetic Structure Supporting the Vertical Threads}
 
The aforementioned comparisons between models and observations suggest that the twisted flux rope model (Model 4) best matches the 
observations, while the sheared arcade model (Model 1) shows the worst match in comparison to observations. 
\citet{2014ApJ...792L..38X} presents a simulation of the in situ condensation process of solar prominences. In this simulation
the vertical prominence resides in the horizontal fields of concave upward field regions of a flux rope. The synthetic SDO/AIA view at 304~\AA~and 211~\AA~in Figure 4 by \citet{2014ApJ...792L..38X} match our observations (Figure 1) very well, but not for the other two channels, i.e., 171~\AA~and 193~\AA. The synthetic SDO/AIA  views show a horn and cavity
structure located above the vertical prominence threads at 193~\AA~and 211~\AA. While in observations, the ``U" structure is shown at 171~\AA. 
This prominence is similar to the tornado-like prominences which are often observed to show rotational motion along the vertical axis  \citep{2012ApJ...752L..22L, 2012ApJ...756L..41S}.
It is unclear how horizontal fields of the flux rope can survive in the presence of rotational motions of the vertical structure.

Model 2 is closest to the ``tangled field model" proposed by \citet{2010ApJ...711..164V}, in which the vertical prominence threads are supported by the
tangled fields in a vertical current sheet below a twisted flux rope, and the ``U" structure corresponds to the central vertical 
current layer in the MHD simulation by \citet{2012ApJ...758...60F}.  Due to limitations in the magneto-frictional method, we are not able
to reproduce the vertical current sheet as suggested in the ``tangled field model", instead we obtain a configuration with newly reconnected
arcade located below an X-point.  Moreover, the magnetic configuration can be strongly distorted by the weight of prominence because the magnetic field 
at polar crown is very weak. Therefore, although Model 2 does not match the observations well, we cannot rule out the ``tangled field model" for the target prominence.
Since unlike the twisted flux rope model,  the ``tangled field model" can explain the rotational motions of tornado-like vertical structure. 

\section{Summaries and Conclusions}

In this work, we study the magnetic structure and dynamics of a tornado-like eruptive polar crown prominence using both observations by SDO and STEREO$\_$B and magnetic field modeling.
Our main findings are summarized below:

(1) STEREO$\_$B observes that the prominence is a very long structure consisting of series of vertical threads. AIA observes the prominence at the south-east limb. Prior to the eruption, the prominence gradually becomes much higher and the cavity becomes much bigger as the Sun rotates from 2012 March 6 to March 11. A Horn-like/U-shape structure appears to go through nearly all the vertical threads from top (171~\AA) to bottom (193~\AA).

(2) The slow rise of the prominence begins around 17:00 UT on March 11, it then evolves to fast rise around 22:57 ($\pm$ 7) UT on March 11, when the height of prominence's leading edge is 97 $\pm$ 5 Mm.  In the AIA  field of view, the maximum velocity  is 110 $\pm$ 5  km $s^{-1}$, and the final acceleration is 49 $\pm$ 5 m s$^{-2}$. Comparing with SOHO/LASCO CME observations, we find that the prominence may have followed our height-time fit (i.e., with increasing acceleration) after leaving AIA field of view until nearly 2.5 R{$_{\odot}$}.

(3)  A post-eruption arcade begins to appear around 23:50 UT on March 11. A newly formed U-shape structure begins to show up at the lower part of the rising prominence threads around 00:10 UT on March 12. Around 00:51 UT, we start to see small bright blobs at the bottom of the rising vertical threads. The speed of the rising blobs is faster than the leading edge and bulk motion of the erupting prominence. The blobs that appear earlier (95 km s$^{-1}$) are slower than those appear later (259 km s$^{-1}$).  These blobs can be explained by the tearing mode instability of a thin current sheet where a series of  magnetic islands are recurrently created during magnetic reconnection.

(4) During the eruption, AIA observes dark ribbons seen in absorption at 171~\AA~corresponding to the bright ribbons shown at 304~\AA. The observational characteristics of dark ribbons at 171~\AA~suggests that they might be caused by the erupting filament material falling back along the newly reconfigured magnetic fields. Dark ribbons are also reported by \citet{Xiao2015}, who interpret it as a void region with smaller magnetic field strength and lower plasma density caused by magnetic field deflection during magnetic reconnection. The co-spatial brightenings at 304~\AA~might be partially caused by heating of the plasma due to the kinetic energy of falling filament material compressing the plasma \citep{2013ApJ...776L..12G}. 

(5) Brightenings at the inner edge of the erupting prominence arcade are observed in all AIA EUV channels during the eruption. These brightenings might be caused by the heating due to energy released from reconnection below the rising prominence.

(6) Using the flux rope insertion method, we construct a series of magnetic field models (sheared arcade model, twisted flux rope model,  and unstable model with HFT), then compare with both SDO and STEREO$\_$B observations. Various observational characteristics appear to be in favor of  the twisted flux rope model. However, the ``tangled field model" cannot be ruled out, since it can explain the rotational motions of tornado-like vertical structure which the twisted flux rope model cannot.

(7) We find that the flux rope supporting the prominence enters the regime of torus instability at the onset of the fast rise phase, and signature of reconnection (post-eruption arcade, new U-shape structure, rising blobs) appears about one hour later. This result suggests that the transition from slow rise to fast rise phase of this prominence eruption is likely to be caused by the torus instability.
 
Acknowledgments: We thank the referee for helpful comments to improve the manuscript. Y.N. Su is grateful to Drs. Cheng Fang, Bernhard Kliem, Xin Cheng, Eric Priest, and Bin Chen for valuable discussions. We thank the team of SDO/AIA, SDO/HMI, STEREO/EUVI for providing the valuable data. The STEREO and HMI data are downloaded via the Virtual Solar Observatory and the Joint Science Operations Center. This project is partially supported under contract SP02H1701R from LMSAL to SAO as well as NASA grant NNX12AI30G.  This work is also supported by the Natural Science Foundation of China (NSFC) No. 11333009, 11173062, 11473071, and the Youth Fund of Jiangsu No. BK20141043.

\newpage

\begin{figure} 
\begin{center}
\epsscale{1.0} \plotone{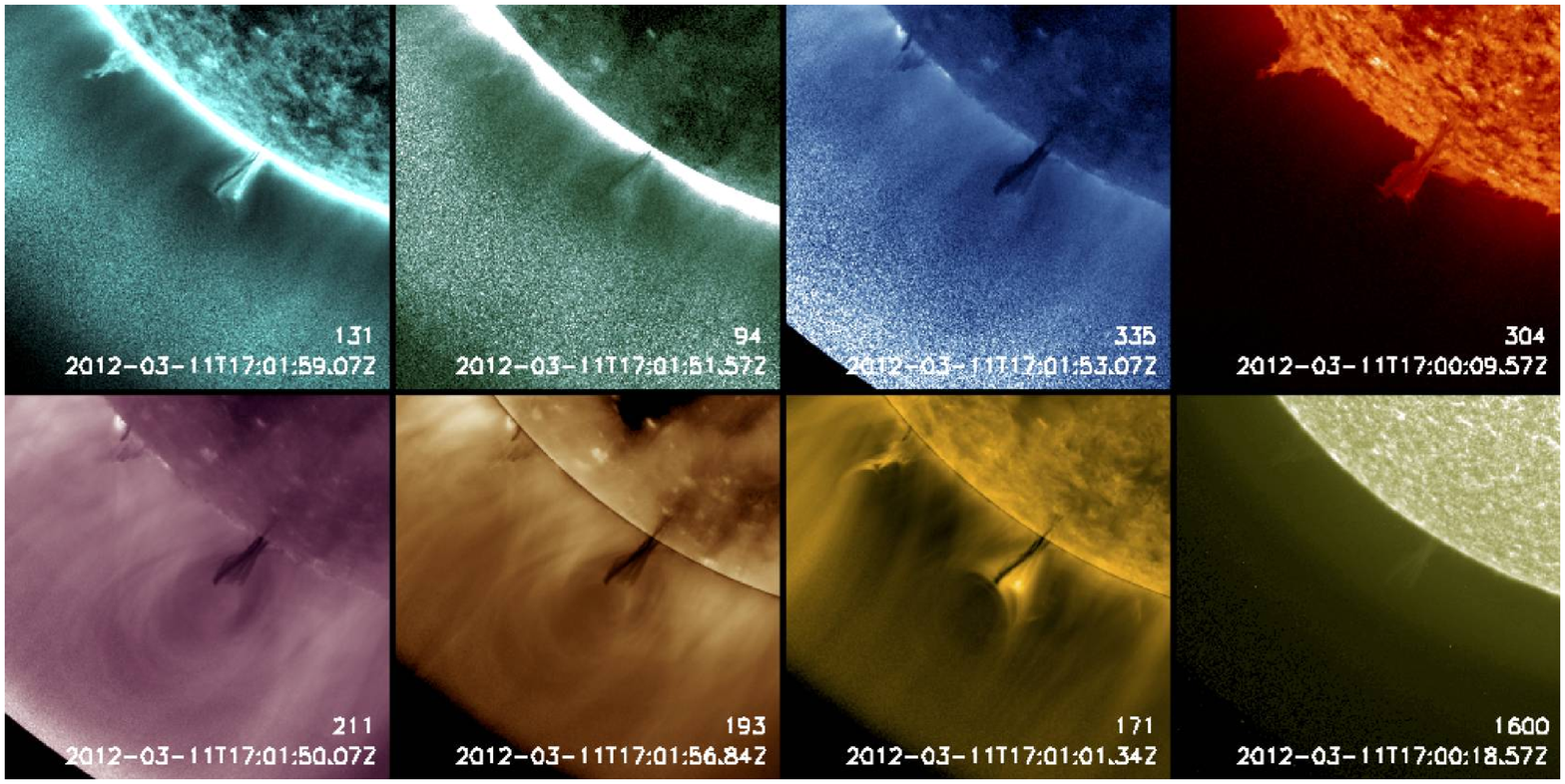}     
\end{center}
\caption {Multi-wavelength images taken by SDO/AIA at 17:00 UT on 2012 Mar 11 before the eruption. A color version of the figure is also available in the electronic edition of the \emph{Astrophysical Journal}. }
\label{fig1}
\end{figure}

\begin{figure} 
\begin{center}
\epsscale{1.0} \plotone{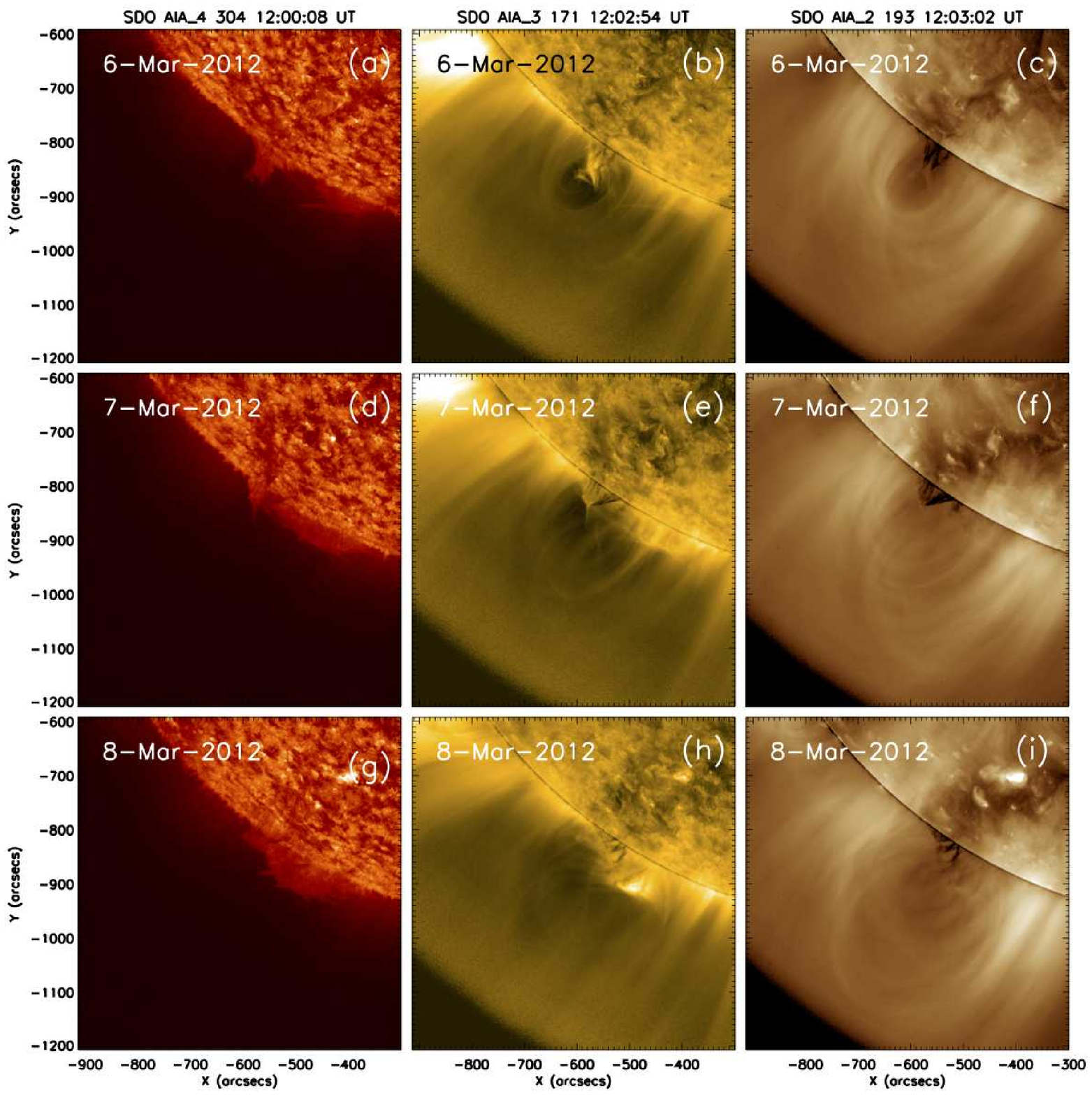}     
\end{center}
\caption {SDO/AIA observations of the eruptive prominence before the eruption at 12:00 UT from Mar 6 to Mar 8 in 2012. A color version of the figure is also available in the electronic edition of the \emph{Astrophysical Journal}.}
\label{fig2}
\end{figure}

\begin{figure} 
\begin{center}
\epsscale{1.0} \plotone{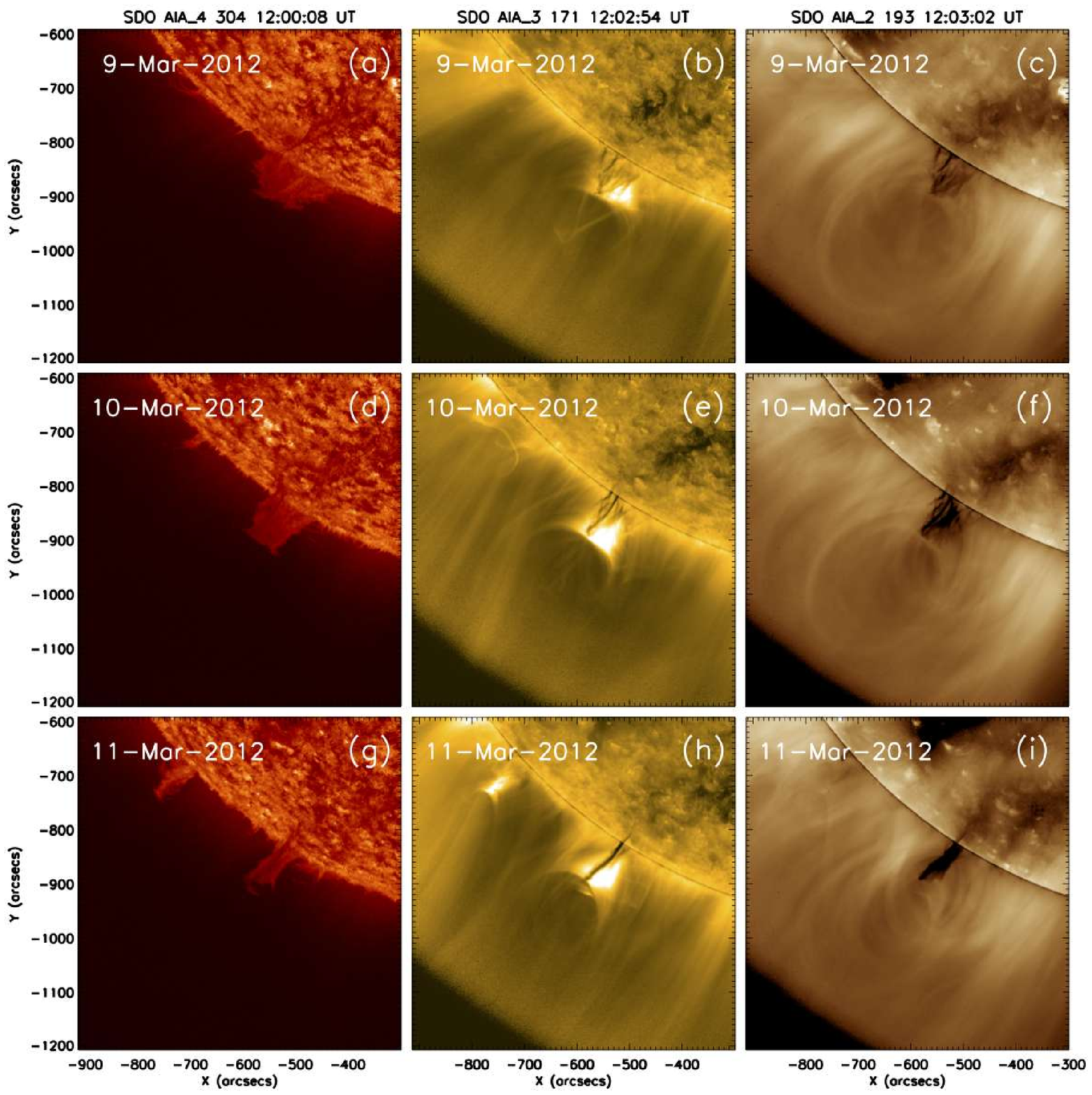}     
\end{center}

\caption{SDO/AIA observations of the eruptive prominence before the eruption from Mar 9 to Mar 11 in 2012. A color version of the figure is also available in the electronic edition of the \emph{Astrophysical Journal}.}
\label{fig3}
\end{figure}

\begin{figure} 
\begin{center}
\epsscale{0.9} \plotone{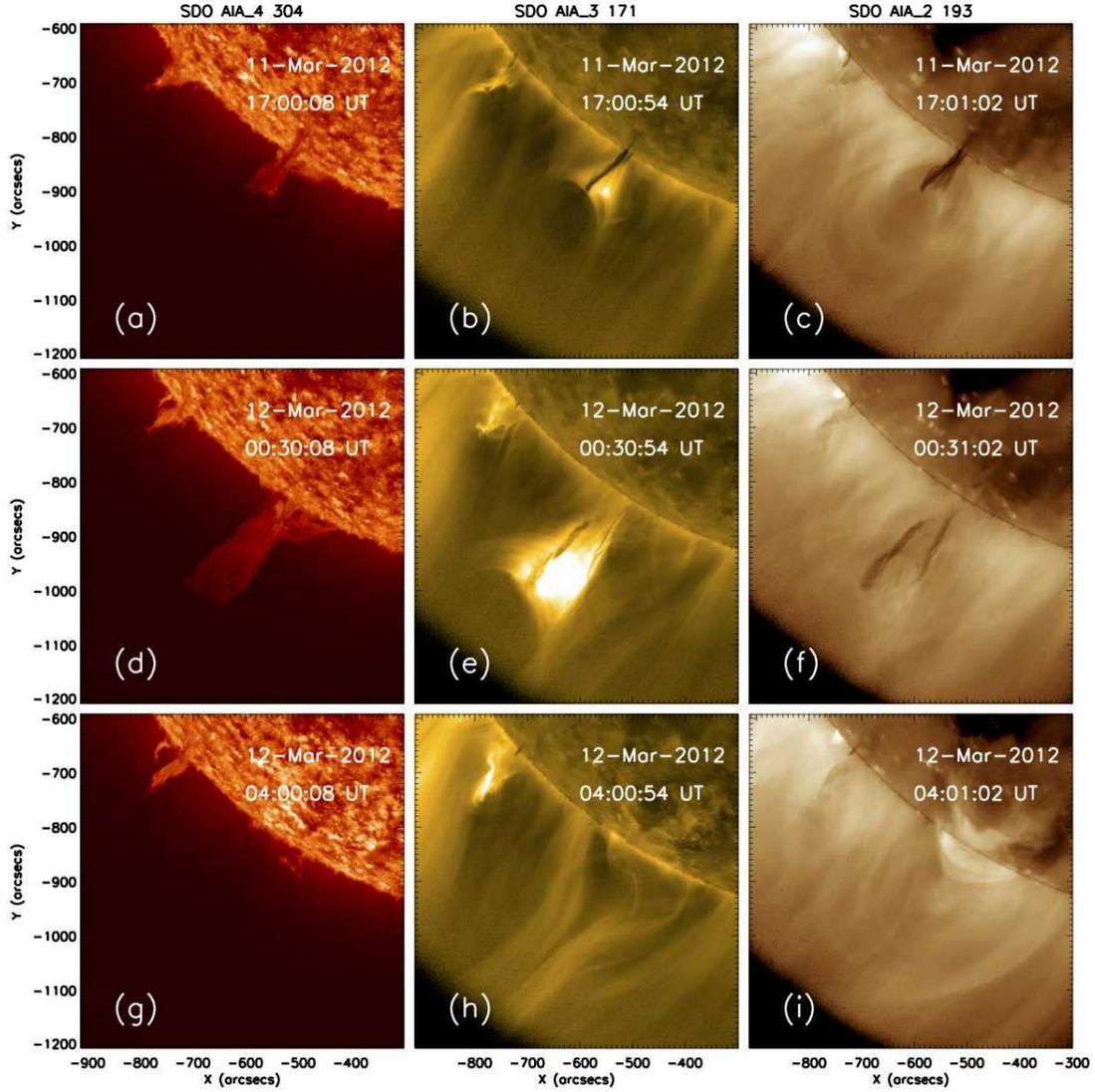}     
\end{center}
\caption{SDO/AIA observations of the prominence eruption on 2012 March 12. A color version of the figure and a video of this event (video 1) are also available in the electronic edition of the \emph{Astrophysical Journal}.}
\label{fig4}
\end{figure}

\begin{figure} 
\begin{center}
\epsscale{0.9} \plotone{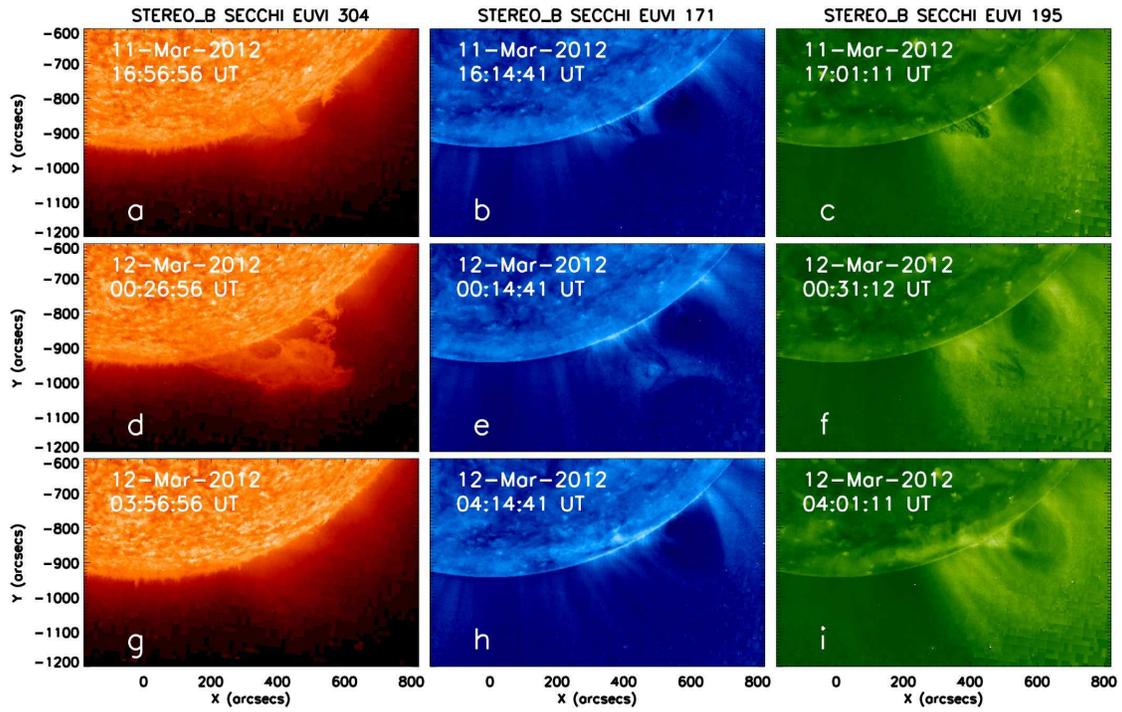}     
\end{center}
\caption{STEREO$\_$B/EUVI observations of the prominence eruption on 2012 March 12. A color version of the figure and a video (video 2) of this event observed by STEREO$\_$B/EUVI are also available in the electronic edition of the \emph{Astrophysical Journal. }}
\label{fig5}
\end{figure}

\begin{figure} 
\begin{center}
\epsscale{0.8} \plotone{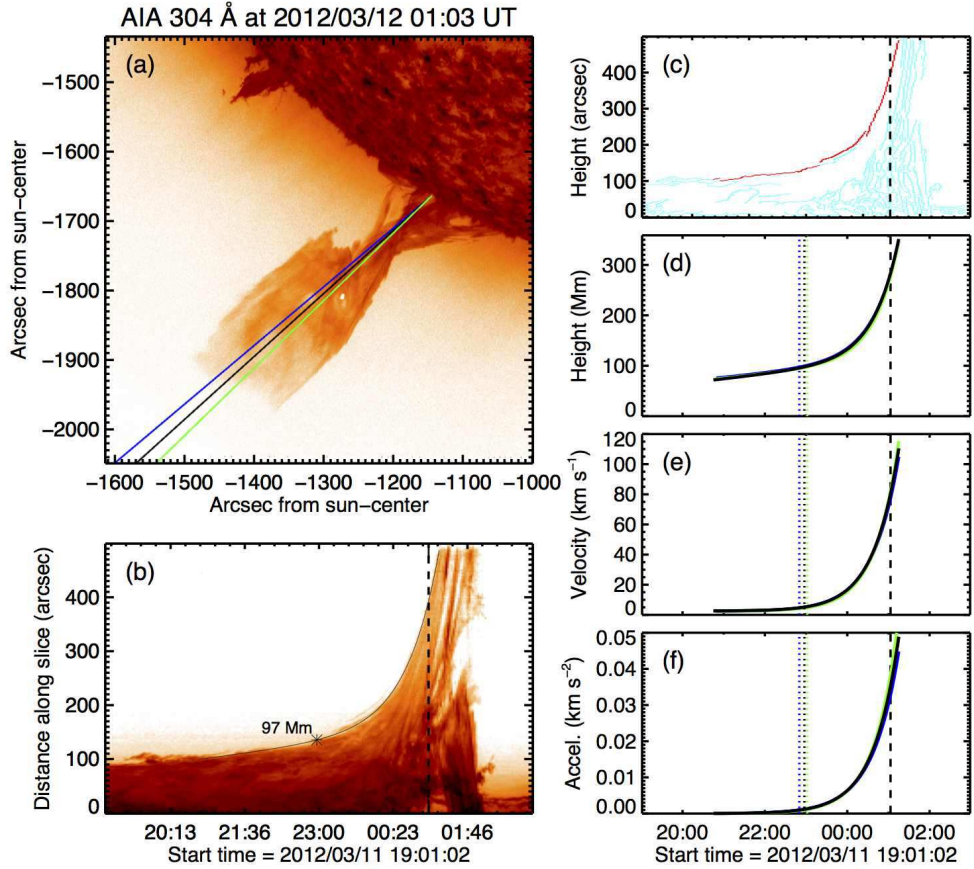}     
\end{center}
\caption{ (a) Trajectories used for tracking. The black slice is selected manually, and the others are offset by 2$^{\circ}$ in either direction.
 (b) Height-time image for the black slice in Panel A. The dashed line indicates the time shown in Panel A, and asterisk indicates the fast-rise onset point. 
 (c) Output of the Canny edge detection algorithm applied to Panel B. The red pixels are used as individual height measurements.  
 (d) Fit to the red pixels in Panel C. The dotted lines indicate the onset of the fast-rise phase, and the colors correspond to results from the different slices in Panel A.
 (e) Velocity and (f) acceleration profiles for the height profile in Panel D. A color version of this figure is also available in the electronic edition of the \emph{Astrophysical Journal}.}
\label{fig6}
\end{figure}

\begin{figure} 
\begin{center}
\epsscale{0.8} \plotone{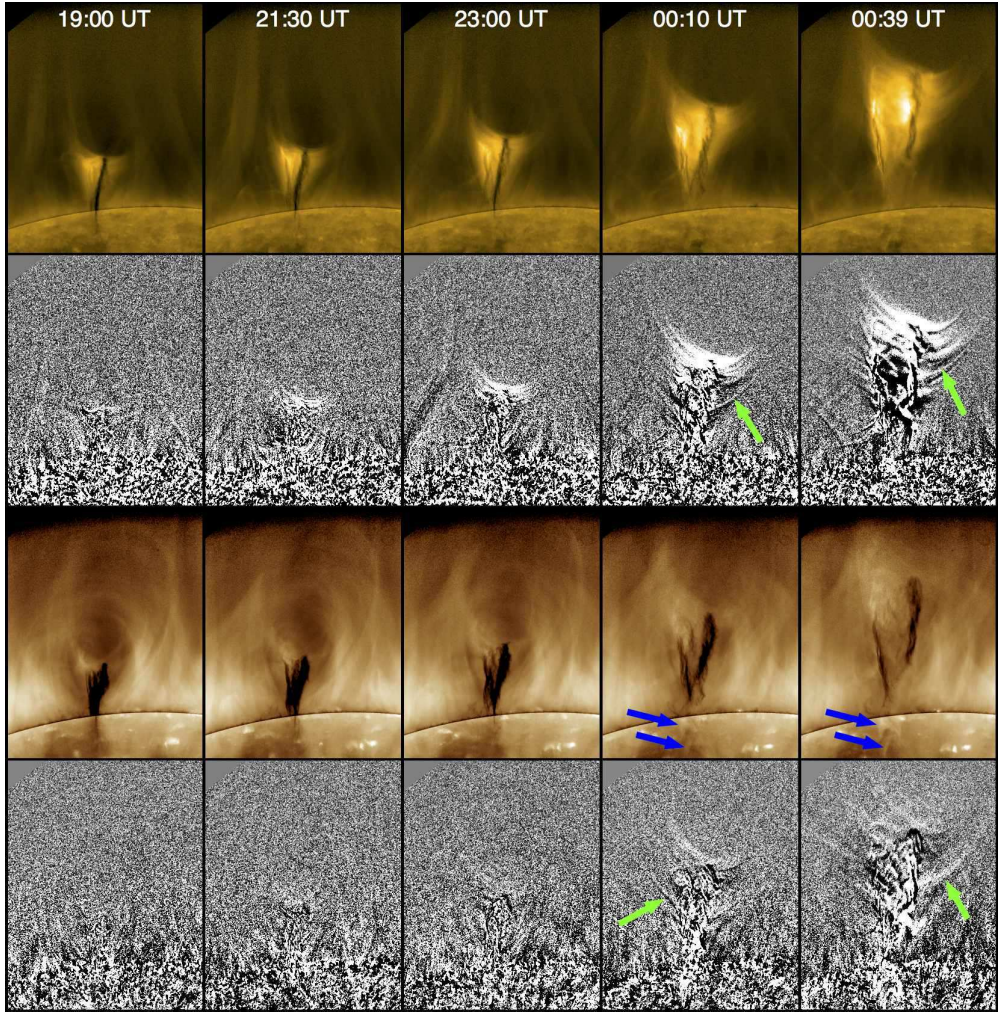}     
\end{center}
\caption {Newly reconnected U-shaped loops (green arrows) and post-eruption arcade (blue arrows). The first and third rows show AIA images at 171~\AA~ and 193~\AA~at different time during the eruption. The images are first summed then radial filter technique is applied to enhance the emission above the limb, and the cadence is 36 seconds. The second and fourth rows show AIA running-difference images at 171~\AA~ and 193~\AA~ at different times during the eruption. The images within 36 seconds are firstly summed then running difference is applied. A color version of this figure and a video (video 3) are also available in the electronic edition of the \emph{Astrophysical Journal}. }
\label{fig7}
\end{figure}

\begin{figure} 
\begin{center}
\epsscale{0.8} \plotone{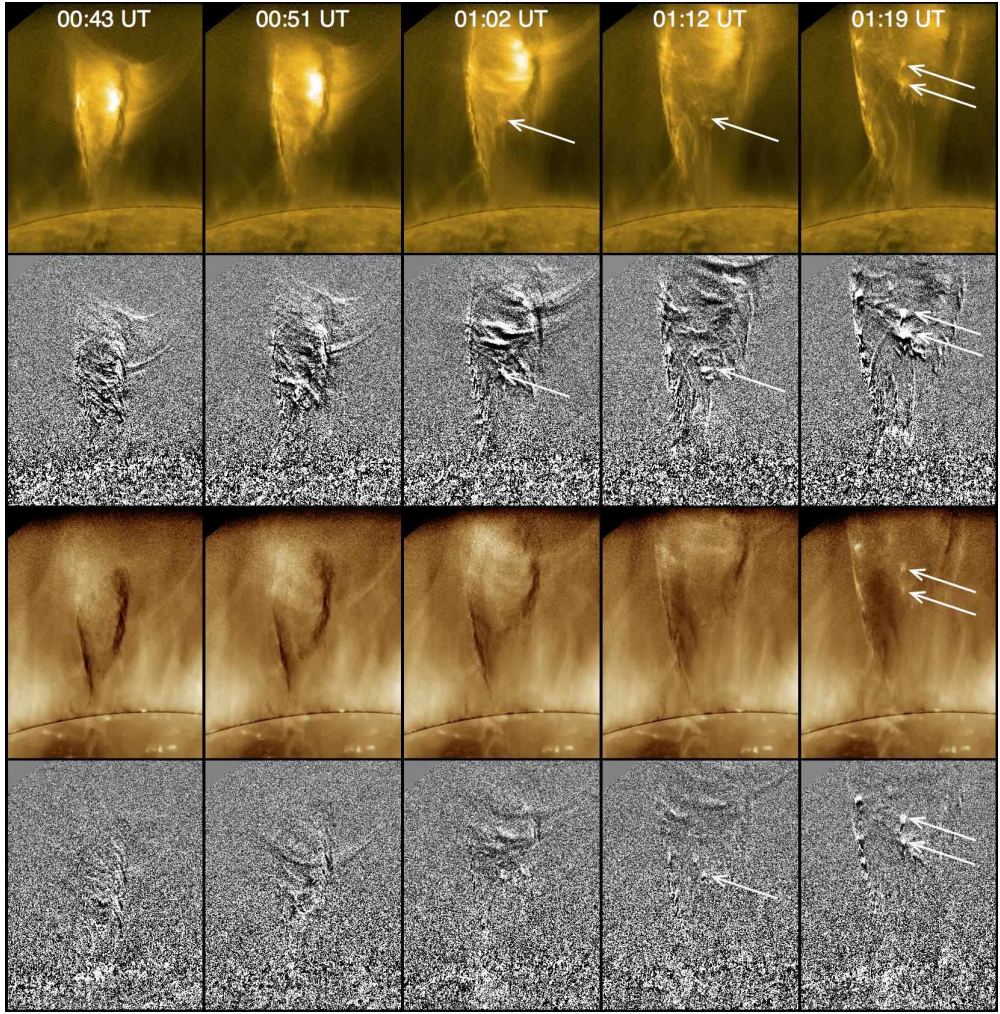}     
\end{center}
\caption{Observations of the rising blobs. The layout of this figure is the same as that in Figure 7. The time of the images in this figure is later than those in Figure 7.  The bright blobs are marked with white arrows. A color version of the figure and a video (video 3) are also available in the electronic edition of the \emph{Astrophysical Journal}.}
\label{fig8}
\end{figure}

\begin{figure} 
\begin{center}
\epsscale{0.8} \plotone{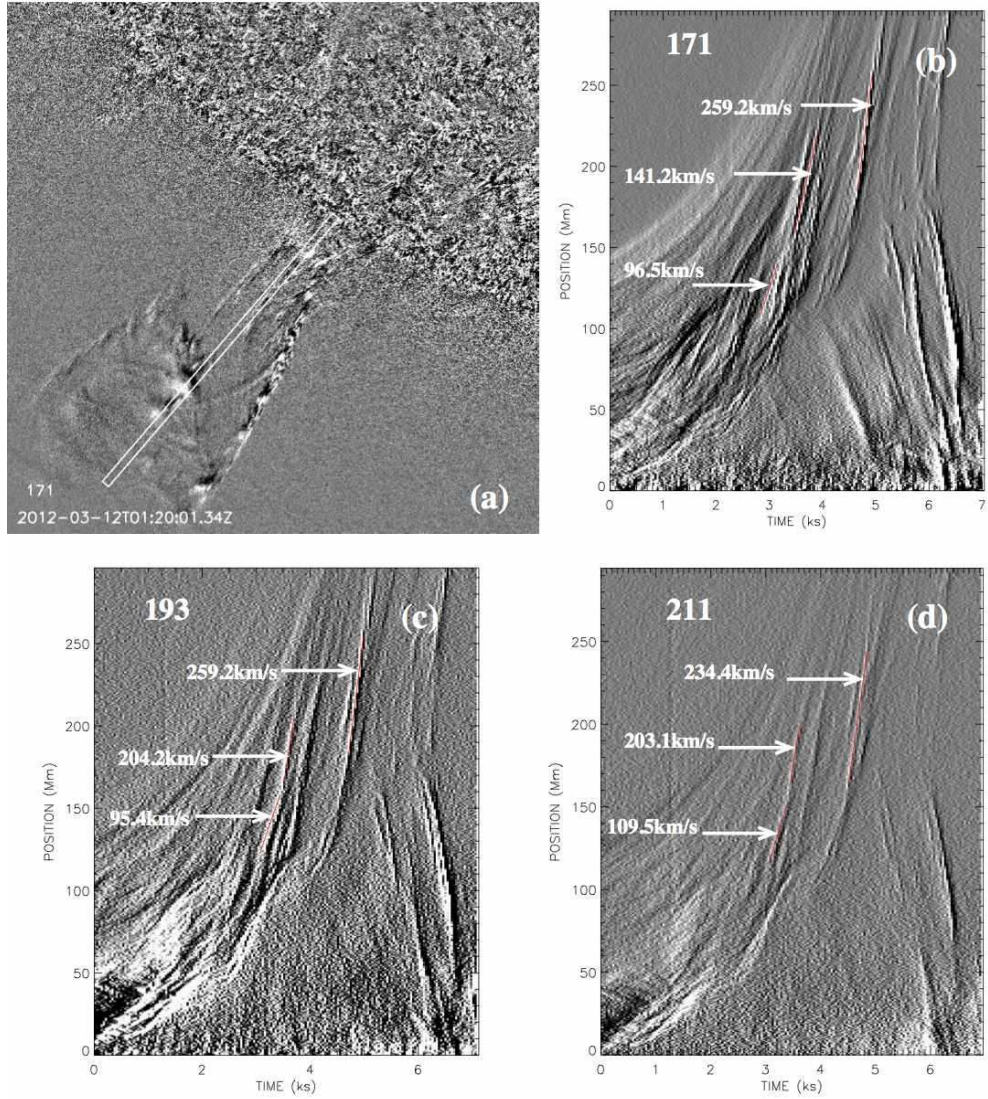}     
\end{center}
\caption{Height-time analysis of the rising blobs from 00:00 UT to 02:00 UT on 2012 March 12. AIA image at 171~\AA~at 01:20 UT on 2012 Mar 12. (b)-(d) refer to position-time plots of emission along the white stripe shown in (a) at 171~\AA, 193~\AA, 211~\AA. Velocities are derived from linear fits to height-time plots of several blobs  marked with red lines. }
\label{fig9}
\end{figure}

\begin{figure} 
\begin{center}
\epsscale{1.0} \plotone{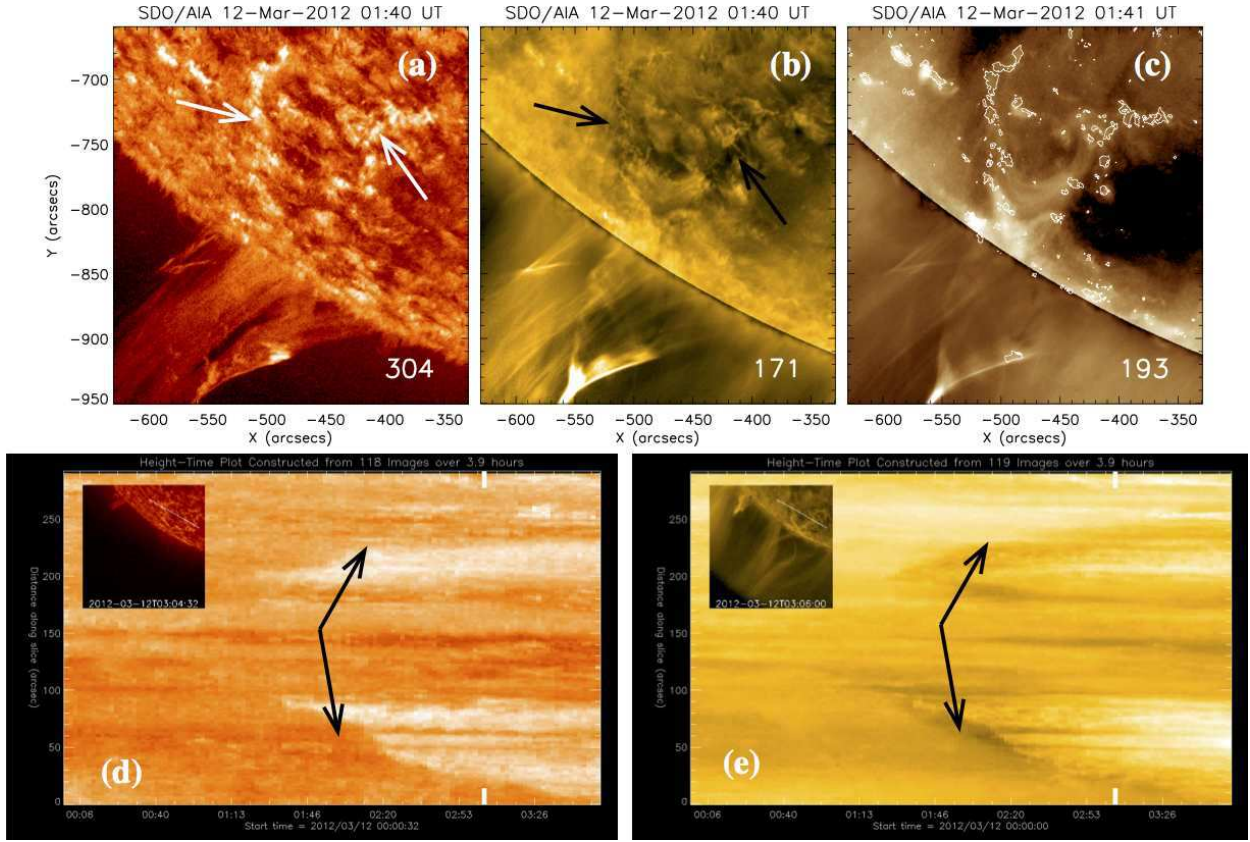}     
\end{center}
\caption{AIA observations of bright (304~\AA, white arrow) and dark (171~\AA, black arrow) ribbons  during the eruption. (a)-(c) AIA images at 01:40-01:41 UT on 2012 March 12 at 304~\AA, 171~\AA, and 193~\AA.  The contours of bright ribbons at 304~\AA~is displayed  on (c). (d) Stack (from 00:00UT and last 3.9 hours ) plot of the emission at 304~\AA~ along the white slice shown in the top left conner of the image. (e) The same as (d) but for 171~\AA~. The black arrows refer to the bright and dark ribbons.}
\label{fig10}
\end{figure}

\begin{figure} 
\begin{center}
\epsscale{1.0} \plotone{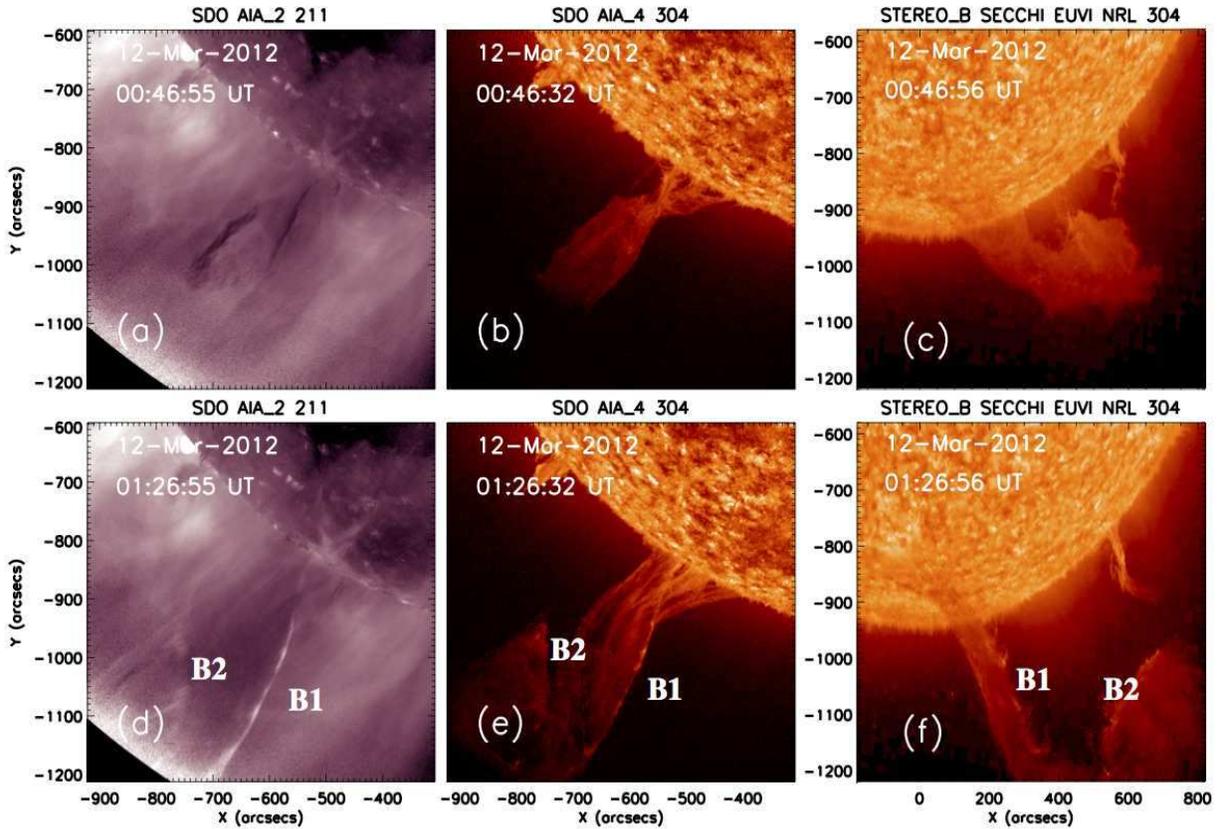}     
\end{center}
\caption{Prominence brightening observed by SDO and STEREO. The first and second columns show AIA images at 211~\AA~and 304~\AA, and  STEREO-B images at 304~\AA~are shown in the third column. The top and bottom rows show images at different time. The prominence brightenings are marked with letter ``B1'' and ``B2". A color version of the figure is also available in the electronic edition of the \emph{Astrophysical Journal}. }
\label{fig11}
\end{figure}

\begin{figure} 
\begin{center}
\epsscale{1.0} \plotone{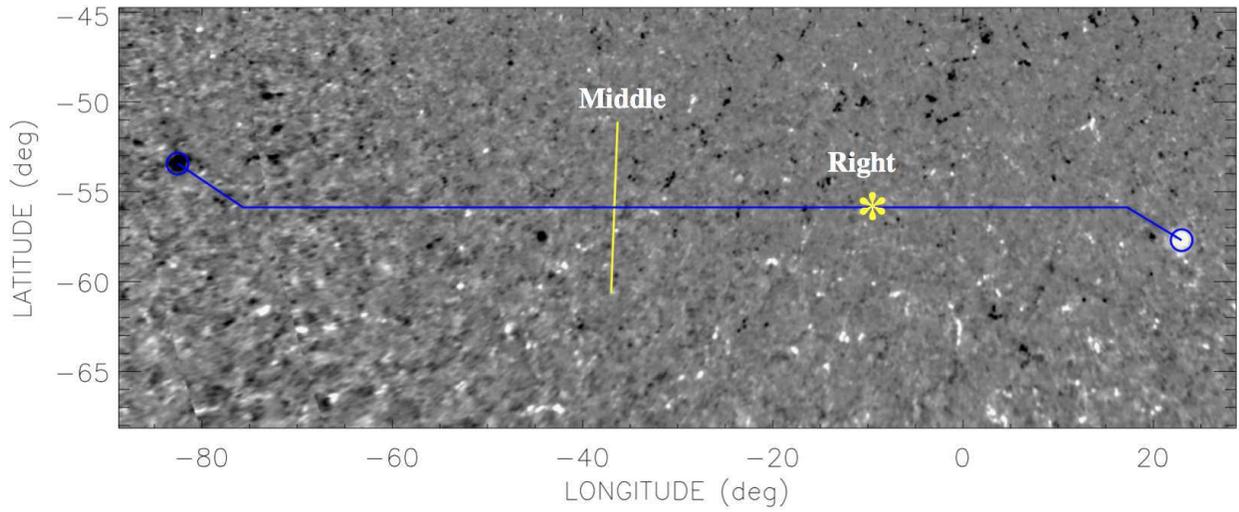}     
\end{center}
\caption{Longitude-latitude map of the radial component of magnetic field in the photosphere in the HIRES region of the model. The longitude zero-point  corresponds to the central meridian on 2012 March 16 at 18:11 UT. The blue curve shows the path along which the flux rope is inserted into the model. The purpose of the yellow line and star signs will be described in the caption of later figures.}
\label{fig12}
\end{figure}

\begin{figure} 
\begin{center}
\epsscale{1.0} \plotone{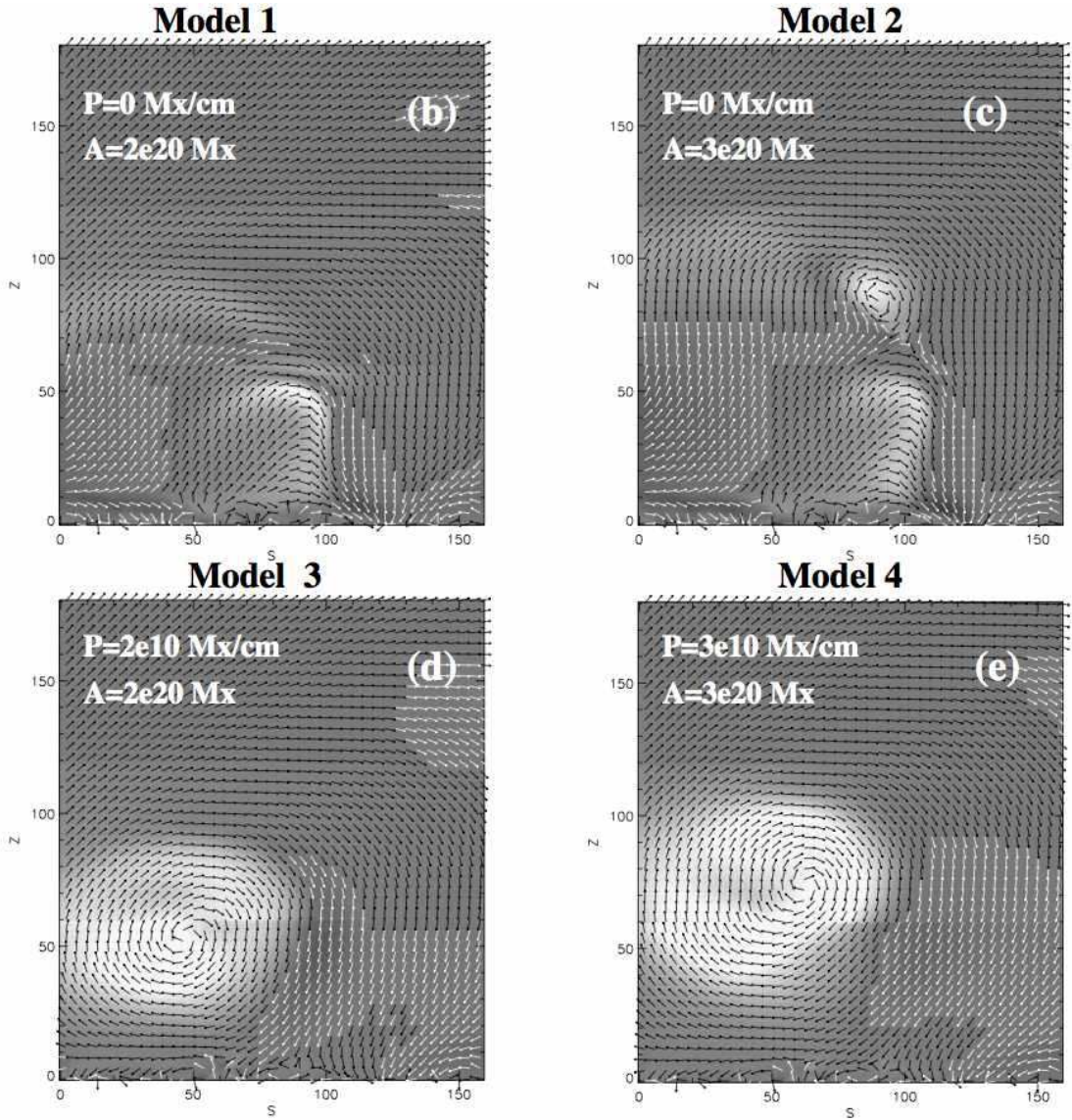}     
\end{center}
\caption{Slices of current distributions along the yellow line in Figure 12 for four different magnetic field models after 80000-iteration relaxations. The initial poloidal and axial flux of the inserted flux is shown in each panel. }
\label{fig13}
\end{figure}

\begin{figure} 
\begin{center}
\epsscale{1.0} \plotone{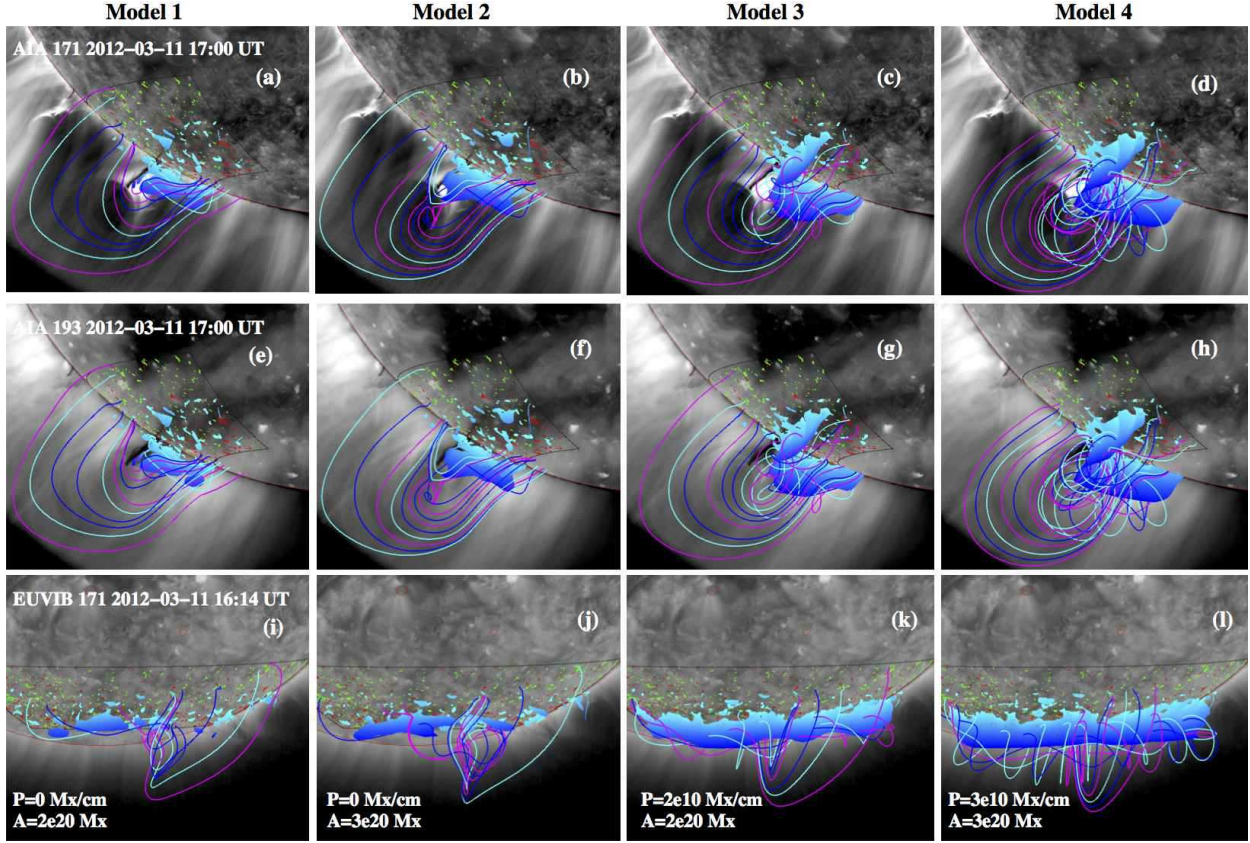}     
\end{center}
\caption{Comparison of the four magnetic field models with observations. The background images at the first, second, and third rows are observed at 171~\AA~ and 193~\AA~ by AIA, and at 171~\AA~by STEREO-B, respectively.  The color curves and blue features refer to selected magnetic field lines and field-line dips from four different models.  The red and green contours show the observed HMI photospheric flux distribution as shown in Figure 12.}
\label{fig14}
\end{figure}

\begin{figure} 
\begin{center}
\epsscale{0.8} \plotone{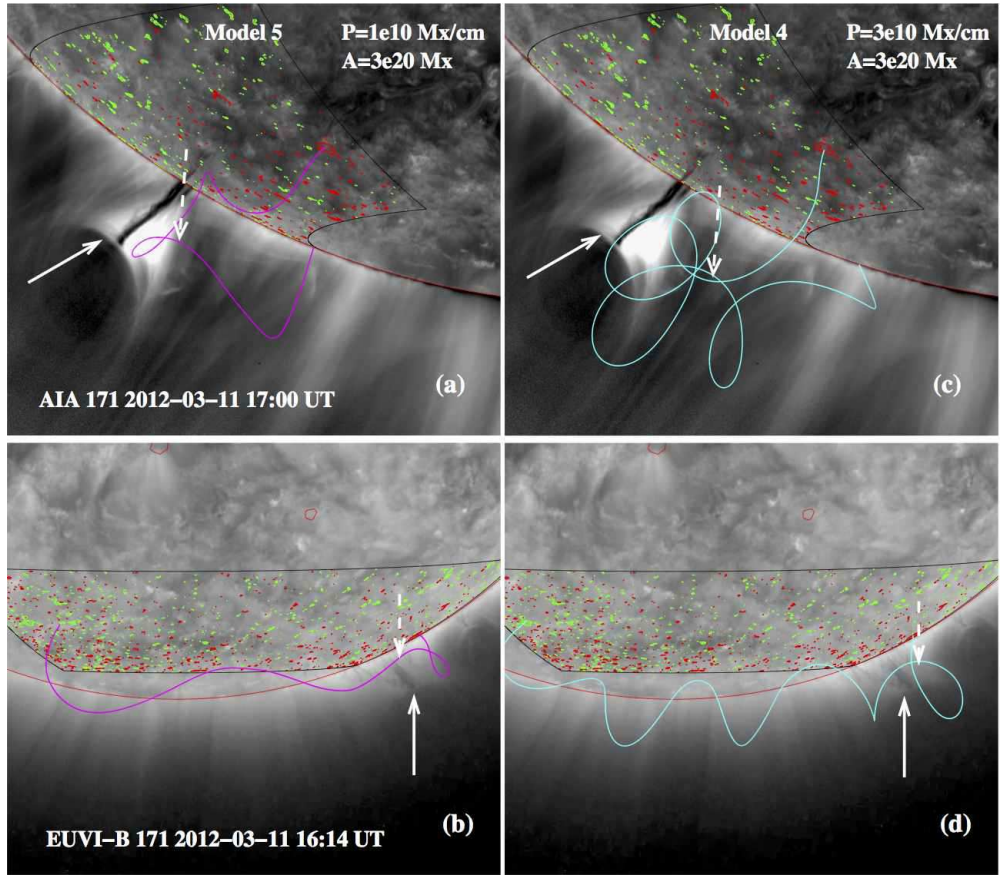}     
\end{center}
\caption{Selected field lines of flux ropes from two different models (left and right) compared with observations.  The background images on the top and bottom panels are at 171~\AA~taken by AIA and STEREO-B before the eruption. The red and green contours show the observed HMI photospheric flux distribution as shown in Figure 12. The U-shape structure in observations are marked with solid white arrow, the curved feature of the model field lines are marked with dashed white arrow.}
\label{fig15}
\end{figure}

\begin{figure} 
\begin{center}
\epsscale{1.0} \plotone{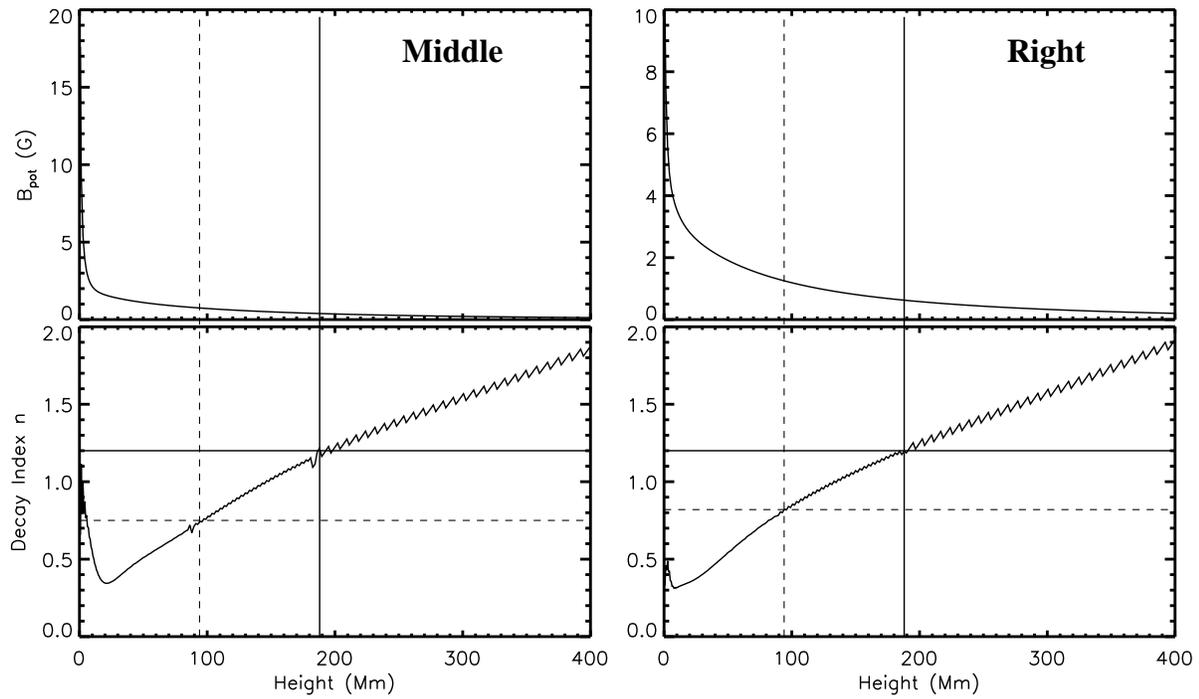}     
\end{center}
\caption{The magnitude of total magnetic field and decay index versus height plots from the potential field model. The left, middle, and right plots are taken at different places which are marked as star signs in Figure 12.}
\label{fig16}
\end{figure}

\end{document}